\documentclass[a4paper,fleqn,usenatbib]{mnras}

\usepackage{txfonts}

\usepackage[T1]{fontenc}
\usepackage{ae,aecompl}
\usepackage{graphicx}	

\title[Kinematics of the Galactic disk]{Kinematics of the Galactic disk from LAMOST Dwarf sample }

\author[Jing et al.]{Yingjie Jing,$^{1}$ Cuihua Du,$^{1}$\thanks{E-mail: ducuihua@ucas.ac.cn} Jiayin Gu,$^{2}$ Yunpeng Jia,$^{3}$ Xiyan Peng,$^{3}$ Yuqin Chen,$^{3}$ \newauthor
Zhenyu Wu,$^{3}$ Jun Ma,$^{3}$ Xu Zhou,$^{3}$ Zihuang Cao,$^{3}$ Yonghui Hou,$^{4}$ Yuefei Wang$^{4}$ \newauthor
and Yong Zhang$^{4}$
\\
$^{1}$School of Physical Sciences, University of the Chinese Academy of Sciences, Beijing 100049, China\\
$^{2}$Department of Physics, Wuhan University of Technology, Wuhan 430000, China\\
$^{3}$Key Laboratory of Optical Astronomy, National Astronomical Observatories, Chinese Academy of Sciences, Beijing 100012, China\\
$^{4}$Nanjing Institute of Astronomical Optics $\&$ Technology, National Astronomical Observatories, Chinese Academy of Sciences, Nanjing 210042, China\\
}

\date{Accepted XXX. Received YYY; in original form ZZZ}

\pubyear{2015}

\begin{document}
\label{firstpage}
\pagerange{\pageref{firstpage}--\pageref{lastpage}}
\maketitle

\begin{abstract}

Based on the LAMOST survey and Sloan Digital Sky Survey (SDSS), we use
low-resolution spectra of 130,043 F/G-type dwarf stars to study the
kinematics and metallicity properties of the Galactic disk.
Our study shows that the stars with poorer metallicity and larger vertical distance from Galactic plane tend to have larger eccentricity and velocity dispersion.
After separating the sample stars into likely thin-disk and thick-disk sub-sample, we find that there
exits a negative gradient of rotation velocity $V_{\phi}$ with metallicity [Fe/H] for the likely thin-disk sub-sample, and the thick-disk sub-sample exhibit
a larger positive gradient of rotation velocity with metallicity. By comparing with model prediction, we consider the radial migration of stars appears to have influenced on the thin-disk formation. In addition, our results shows that the observed thick-disk stellar orbital eccentricity distribution peaks at low eccentricity ($e \sim 0.2$) and extends to a high eccentricity ($e \sim 0.8$).  We compare this result with four thick-disk formation simulated models, and it imply that our result is consistent with gas-rich merger model.

\end{abstract}


\section{Introduction}

The formation and evolution of the Galaxy are very
important issues in modern astrophysics. Since
\citet{Gilmore83} firstly introduced the thick-disk,
the basic components of the Galactic disk are
thin-disk and thick-disk populations.
The two components differ not only in their spatial
distribution profiles but also in their kinematics and metallicity
\citep{majewski93,Ojha96,Freeman02}.
In the spatial distribution, the range of scale height
for the thin-disk vary from 220 to 320 pc, while that of the thick-disk is from 600 to 1100 pc \citep{Du03,Du06,Jia14}.
The range of scale length for the disk (including thin disk and thick disk) vary from 2 to 4 kpc,
but some evidences for a short scale length for the thick disk were also given by  \citep[e.g.][]{Bensby11,Cheng12,Hayden15}.
The thick-disk generally has lower rotational velocity and
 larger velocity dispersion, has a lower average metallicity
 ([Fe/H] $\sim -0.7$ dex)\citep{Gilmore85}, and has enhanced
 $\alpha$-element abundances than the thin disk
 \citep{Gratton96, Fuhrmann98, Prochaska00, Bensby03,
Reddy06, Bensby07, Fuhrmann08}.

The origin of the
thick disk have been investigate by authors \citep[e.g.][]{Quinn86,
Freeman87, Abadi03, Brook04, Brook05, Brook07, Schonrich09} and have not
been resolved. Some currently discussed models of formation
mechanisms for the thick disk predict various trend between the
kinematics properties and metallicity of disk stars, as well as
between their kinematics and spatial distributions. For example,
models of Gas-rich merger predict a rotational velocity gradient
with Galactocentric distance for disks stars near the solar radius
\citep[][]{Brook07}.  Models of disk heating via satellite
mergers or a growing thin disk can induce a notable increase in the
mean rotation and velocity dispersions of thick disk stars
\citep[][]{Villalobos10}. \citet{Sales09} also showed that the
distribution of orbital eccentricities for nearby thick disk stars
could provide constraints on these proposed formation models.
Comparisons the predictions of the models with observed kinematics properties of Galactic disk are helpful to understand the formation and evolution of Galaxy.

To understanding the formation and chemical evolution of the Galaxy components,
we need more chemical and kinematics information of a large number of stars in larger areas which will
greatly increase the spatial coverage of the Galaxy.
The large-scale spectroscopic surveys make it possible by providing ideal databases such as
radial velocities and stellar atmospheric parameters (Teff, log $g$, [Fe/H], etc).
A number of papers have employed the kinematics
characters and chemical abundances to study the Galactic structure
and formation, based on spectroscopic survey data. For instance,
\citet{Bond10}, \citet{Carollo10} , \citet{Lee11} and
\citet{Smith09} have characterized the halo and disk base
on Sloan Digital Sky Survey \citep[SDSS;][]{York00} and its
sub-survey Sloan Extension for Galactic Understanding and
Exploration \citep[SEGUE;][]{Aihara11,Yanny09, Beers06}.
SDSS III's Apache Point Observatory Galactic
Evolution Experiment \citep[APOGEE;][]{Eisenstein11} has higher resolution than SEGUE, also being used to explore the kinematic of disk \citep[e.g.][]{Bovy12,Bovy15, Ness16}.
Comparing to SEGUE,  the Radial Velocity Experiment (RAVE) provide a bright
complement to the SEGUE sample \citep{Siebert11}. Many works on
kinematics of Galactic disk
\citep[e.g.][]{Binney14,CasettiDinescu11,Siebert08} have been done on RAVE
data.
Of course, there also have some works based on Gaia-ESO internal data-release \citep[e.g.][]{RecioBlanco14,Kordopatis15}.
The ongoing Large Sky Area Multi-Object
Fiber Spectroscopic Telescope survey \citep[LAMOST, also called Guoshoujing
Telescope;][]{Cui12, Deng12, Zhao12, Luo12} has release more than two
millions stellar spectra with stellar parameters in the DR2 catalog.
This data set will provide a vast resource to study details of the
velocity distribution and give constraints on the dynamical
structure and evolution of the Galactic disk.

In this study, we make use of the F/G dwarf stars selected from the LAMOST survey to
explore the observed correlations of kinematic velocity and orbital
eccentricity with metallicity and distance from the Galactic plane.
We also compare the observation results with the predictions
 of different models and expect to obtain some clues for the disk formation.
The outlines of this paper are as follows. In Section 2,
we take a brief overview of the LAMOST, observation data and derive the individual three-dimensions velocity.
Section 3 presents the results of the observation. The discussions are given in
Section 4. A summary and conclusions are given in Section 5.

\section{The data}

\subsection{Observational Data}
The Large Sky Area Multi-object Fiber Spectroscopic Telescope (LAMOST)
survey is a 4 meter quasi-meridian reflective Schmidt telescope at
the Xinglong Station of the National Astronomical Observatories
(NAOC), Chinese Academy of Sciences. The field of view of the CCD is
$5^{\circ}$. It is a powerful instrument to survey the sky with the
capability of recording 4,000 spectra simultaneously.  The
spectrograph of LAMOST has a resolution of R $\rm \sim$ 1,800 and
range spanning 3,700{\AA} $\sim$ 9,000{\AA} \citep{Wang96, Su04, Cui12}.
LAMOST has completed 3 years of survey operations plus a Pilot Survey, and internally
released a total $\sim 5.7$ million spectra to the collaboration.
Of these, $\sim 3.7$ million are AFGK-type stars with
estimated stellar atmospheric parameters as well as radial
velocities.  Stellar atmospheric parameters (including spectral
types, [Fe/H], log $g$) are derived by Ulyss software \citep{Wu11}.
The radial velocities are measured by cross-correlation between
spectra and template spectra from the Elodie library \citep{Moultaka04}.
The survey reaches a limiting magnitude of $r=17.8$ (Where $r$ denotes magnitude in the SDSS $r$-band),
and most targets is brighter than $r\sim17$.

In this study, we use stellar atmospheric parameters (including
spectral types, [Fe/H], log $g$) and radial velocity from LAMOST DR2
catalogs.  Our initial sample was obtained by a cross referencing
between the LAMOST DR2 and SDSS DR12 photometric catalog based on
stellar position. The proper motions of the sample are obtained by
cross-identifying with SDSS-POSS proper-motion catalog
\citep{Munn04, Munn08}. Stars without SDSS photometry and proper
motions were eliminated to estimate photometric distances by SDSS
color.  Next, we select F/G dwarf stars (log $g > 3.5$) with $\rm
S/N >$ 15 at $g$ band.  In order to use the photometric parallax
relation, we restrict sample at $0.2<g-i<4.0$. This cut reduces the
numbers of stars to  180,759.

\begin{figure}
\includegraphics[width=1.0\hsize]{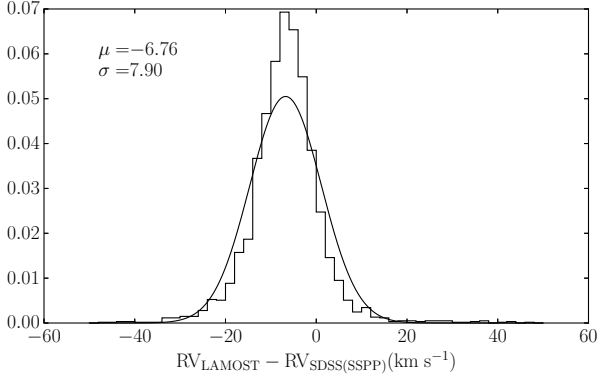}
\caption{The offsets of the radial velocities between the LAMOST
pipeline and the SEGUE Stellar Parameter Pipeline. The histogram is
the offset distribution of 3077 common stars, and the solid curve is
Gaussian fit to the offset distribution. }
\label{figure1}
\end{figure}

In order to investigate the
systematics of the radial velocities, we cross identify the LAMOST
data ($\rm S/N > 20 $ at $g$ band and log $g > 3.5$) with the
SDSS-SSPP data (average $\rm S/N > 20 $) and select 3077 common
stars as a sample with good parameter estimates.  As shown in Figure
\ref{figure1}, the offset of radial velocities between the LAMOST
pipeline and the SDSS-SSPP is concentrate on $-6.76$ km~s$^{-1}$
with a dispersion of 7.9 km~s$^{-1}$. \citet{Tian15} also reports
that the radial velocity derived from the LAMOST pipeline is slower
by $-5.7$ km~s$^{-1}$  compared with APOGEE.  The reason for this
offset is unclear. In this study, in order to match the other survey
data, we add an additional 6.76 km~s$^{-1}$ to the derived LAMOST
radial velocity.

\subsection{The photometric parallax and spatial velocity}

We estimate the distance using a photometric parallax method described
by \citet{Ivezic08}, which gives the absolute magnitude in the $r$ band as
a function of $g-i$ and [Fe/H]. We then derive the $U, V, W$ space velocity
components from the distances, radical velocities and proper
motions. Here, we apply $(U^{\odot,{\rm pec}}, V^{\odot,{\rm pec}}, W^{\odot,{\rm pec}})
= (-11.1, -12.24, 7.25)$ km s$^{-1}$ \citep{Schonrich10} to adjust for the
solar peculiar motions with respect to the local standard of rest (LSR). In order to obtain the Galactocentric cylindrical velocity components, we adopt $V_{\rm LSR} = 220$ km~s$^{-1}$ \citep{Gunn79, Feast97} and $R_{\sun} = 8$ kpc.

\begin{figure}
\includegraphics[width=1.0\hsize]{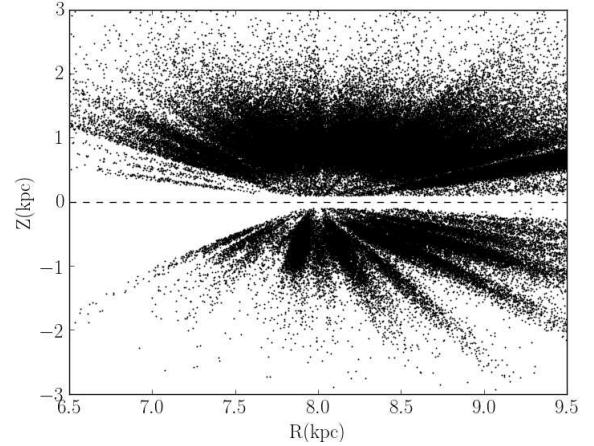}
\caption{Spatial distribution in the the cylindrical Galactic
coordinates $Z-R$ plane of the 130,043 sample stars in this study.
The dashed line represents the Galactic plane. }
\label{figure2}
\end{figure}

\begin{figure}
\includegraphics[width=1.0\hsize]{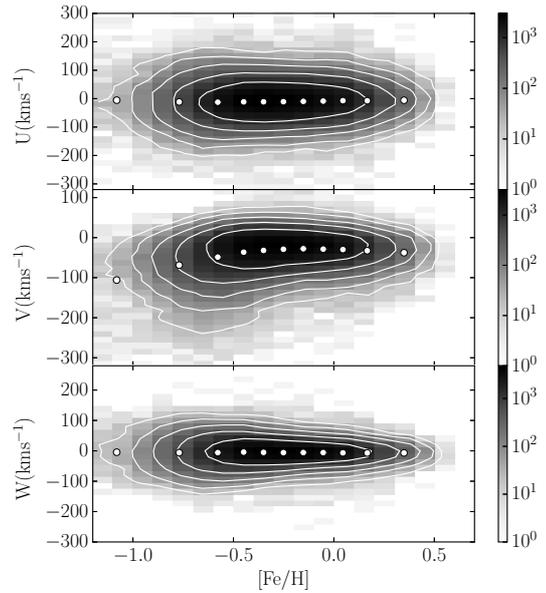}
\caption{The iso-density contours of $U$, $V$ and $W$ distribution
with metallicity [Fe/H], respectively. The filled circles denote the mean value
of $U$, $V$ and $W$ in different metallicity intervals.}
\label{figure3}
\end{figure}

We select these stars with $6.5<R<9.5$ kpc and $0.1<|$Z$|<
3$ kpc, this cut reduces the sample stars to 130,043.
The reasons for the selection criteria $|Z|> 0.1$ kpc are that these stars are not well observed in the LAMOST survey and SDSS survey
because most of them are too bright \citep{Tian15} and to avoid the effects of overestimated interstellar extinction correction for the these stars \citep{Juric08}. The median estimated errors on the
$V_{\phi}, V_{R}$ and $V_{Z}$ of the sample stars are 16, 16 and 17 km s$^{-1}$, respectively.
The spatial distribution of the sample in the cylindrical Galactic coordinates
Z$-R$ panel is plotted in Figure \ref{figure2}.
Figure \ref{figure3} gives the iso-density contours of the velocity
components $U$, $V$ and $W$ of the sample stars with metallicity, and
the filled circles denote the mean value of velocity components in
different metallicity interval. As shown in Figure \ref{figure3}, there is no
significant association with [Fe/H] for $U$ and $W$,
however, for $V$ which is associate with $V_{\phi}$, there has a relative small-scale spatial distribution.
Summarizing the criteria adopted in our sample selection, the sample stars satisfy
$6.5<R< 9.5 \rm ~kpc$, $0.1<|Z|< 3\rm~kpc$, log $g > 3.5$,
$-1.2<$[Fe/H]$<0.6$. It shows that our sample stars could be mainly
contributed from the disk system. In total, there are 130,043 sample stars in this study.

\section{Results of the observations}
\subsection{Correlation of orbital eccentricities with metallicity}

\begin{figure}
\includegraphics[width=1.0\hsize]{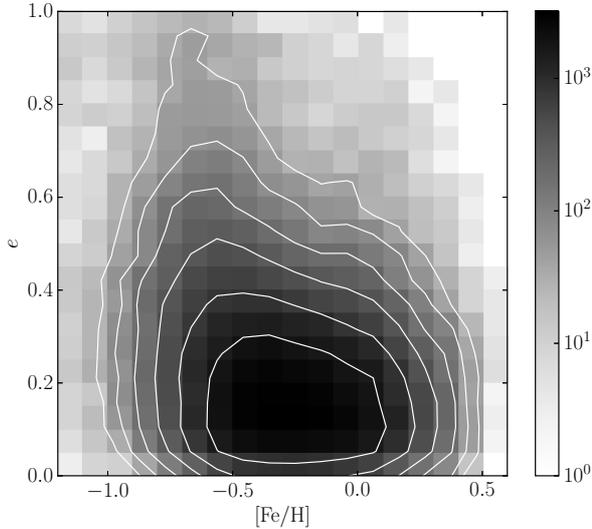}
\caption{Metallicity [Fe/H] vs. orbital eccentricity for the sample
stars.}
\label{figure4}
\end{figure}

We investigate the orbital properties of our sample stars by
adopting the Galaxy potential model of \citet{Paczynski90}
proposed by \citet{Miyamoto75}. The model is axis-symmetric and fully
analytic, and consists of a spherical central bulge, a highly
flattened disk and a nearly spherical halo. We derive the orbital
eccentricity $e$, defined as $e=(r_{\rm apo} - r_{\rm peri})/(r_{\rm
apo} + r_{\rm peri})$, where $r_{\rm peri}$ denote the closest
approach of an orbit to the Galactic center (i.e., the perigalactic
distance), and ${r_{\rm apo}}$ denote the farthest extent of an
orbit from the Galactic center (i.e., the apogalactic distance).
We performed a Monte Carlo simulation to estimate the errors of orbital eccentricities:
for each star,  we computed 1000 times of the orbital eccentricities by changing
the Galactic coordinates and the velocities every time, assuming Gaussian distributions around their values
and a dispersion according to their estimated errors.
From these 1000 orbits we computed the standard deviation for each eccentricity.
The errors of orbital eccentricity are smaller than 0.15 dex for most stars.
Figure \ref{figure4} shows the eccentricity versus metallicity [Fe/H] for the sample stars in this study.
The eccentricity and metallicity of our sample have a wide dispersion. But most
of the stars have an eccentricity $<$ 0.6.

\begin{figure}
\includegraphics[width=1.0\hsize]{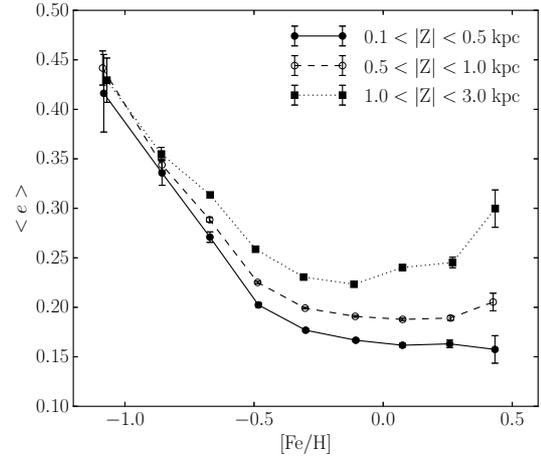}
\caption{The orbital eccentricity as a function of [Fe/H] for
the selected sample. The filled circles, open circles and filled
squares correspond the stars at $0.1< |Z| <$ 0.5 kpc, 0.5 $< |Z| <$ 1
kpc, and 1 $<|Z|<$ 3 kpc, respectively. The bin represent a range
of 0.2 dex in metallicity.} \label{figure5}
\end{figure}

We compute the mean orbital eccentricity in different metallicity
and $|Z|$ distance interval and the results are indicated in Figure
\ref{figure5}. The bin of [Fe/H] is 0.2 dex and the filled
circles, open circles and filled squares denote the stars at $0.1<|Z| <$
0.5 kpc, 0.5 $<|Z|<$ 1 kpc, and 1 $< |Z| <$ 3 kpc, respectively. As
shown in this figure, the eccentricity increases with increasing
distance from the Galactic plane and the change trend of eccentricity versus [Fe/H] is different at $-1.2<$
[Fe/H]$<-0.5$ and $-0.2 <$[Fe/H]$< 0.6$. In the first metallicity
range, the sample is dominated by the
thick-disk population, the trend of the orbital eccentricity
generally decreases as the metallicity increases, and there is a
 relatively larger gradient with increasing metallicity.  In the
second metallicity range which is dominated by
thin-disk population, there is an almost flat trend of orbital
eccentricity with metallicity [Fe/H] (the gradient approach zero
except at 1 $< |Z| <$ 3 kpc), and indicate that the
orbital eccentricity is independent of metallicity. These results
agree well with the recent determinations from the SEGUE data
\citep{Lee11} which also found the thick-disk stars exhibit a strong
trend of eccentricity with metallicity (about $-0.2$ dex$^{-1}$),
while the eccentricity does not change with metallicity for the
thin-disk sub-sample.  However, for those stars in distance interval
1 $<|Z|<$ 3 kpc and the second metallicity range ($-0.2 <$[Fe/H]$<
0.6$),  the eccentricity has a little increase with increasing
metallicity. We consider that these stars are assimilated into the
thick disk from orbits near the Galactic disk plane or radial migration have
influenced the structural and chemical evolution of the Galactic disk. Even, it is
possible that these high eccentric orbit and metal-rich stars
origin from bulge, \citet{Pompeia02} show that there exist bulge-like stars with rich metallicities and very eccentric orbits ($e>0.25$). But their sample stars have nearly distance from the Galactic plane ($|Z|<1$).

\subsection{The properties of rotational velocity}

\begin{figure}
\includegraphics[width=1.0\hsize]{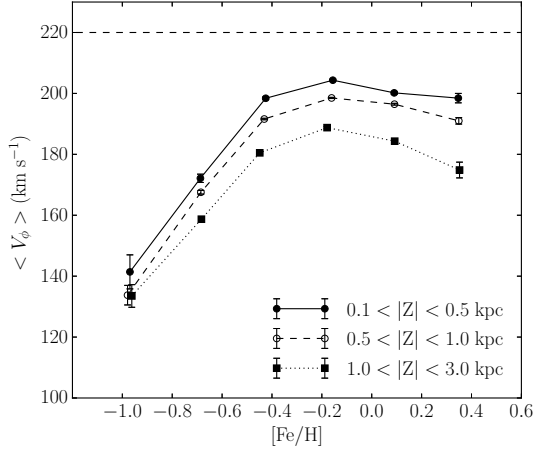}
\caption{Distribution of the mean rotational velocity as a function
of metallicity [Fe/H] for the selected sample with different $|Z|$
intervals. The filled circles, open circles and filled squares
denote the stars at $0.1< |Z|<$ 0.5 kpc, 0.5 $<|Z|<$ 1 kpc, and 1 $< |Z|
<$ 3 kpc, respectively. The dashed line is the adopted $V_{\rm LSR}
= 220$ km~s$^{-1}$ for the motion of the local standard of rest.}
\label{figure6}
\end{figure}

In this section, we use our sample to examine the properties of the
rotational velocity with metallicity [Fe/H], $R$ and $|Z|$. Figure \ref{figure6}
shows the mean rotational velocity distribution as a function of
metallicity [Fe/H] for the selected sample. The metallicity interval and the
value in each interval are also listed in Table \ref{table1}. As
shown in Figure \ref{figure6}, the mean rotational velocity
increases with increased metallicity  and has a peak at
[Fe/H] $\sim -0.2$, then it becomes a slightly decrease when
[Fe/H] $>-0.2$.
A plausible explanation is that high-metallicity
range is dominated by thin-disk population and the rotational
velocity of thin-disk decreases with increased metallicity [Fe/H],
however most stars at lower metallicity belong to thick disk
population and the rotational velocity of thick-disk increases with
increasing metallicity [Fe/H]. In order to check this explanation,
we select the likely thin-disk and thick-disk sub-sample from our
sample following the prescription of \citet{Bensby03, Bensby14}. We
calculate the likelihood each star belongs to the thin-disk,
the thick-disk and the halo assuming the space velocities of
three components are distributed as Gaussian. The adopted value of the local
stellar densities, velocity dispersions in $U$, $V$, and $W$, and
the asymmetric drifts here are the same as that in \citet{Bensby14}.
We select the likely thick-disk stars as those
with $-1.2<$ [Fe/H] $<-0.3$, relative likelihood for the thick-disk
to thin-disk (TD/D) $>$ 5 and thick-disk to halo (TD/H) $>$ 2, the
likely thin-disk stars as those with [Fe/H] $>-0.5$ and TD/D $<$
0.2.

\begin{figure}
\includegraphics[width=1.0\hsize]{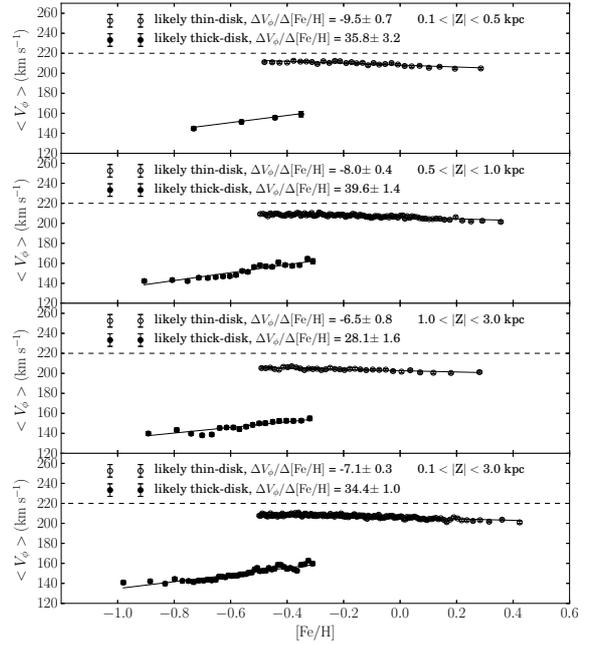}
\caption{Rotation velocity gradients with metallicity [Fe/H] for
different intervals in distance from the Galactic plane, for stars
assigned to likely thin-disk (open circles) and thick-disk (filled
circles) populations. } \label{figure7}
\end{figure}

Figure \ref{figure7} shows the result of this verification. The
values of each circle is obtained by passing a box of 500 stars,
with an overlap of 100 stars per bin through the data. The slopes
are obtained by performing least-square fits to each unbinned sub-sample.
The uncertainties of the slopes are calculated by resampling each sub-sample with replacement 1000 times,
assuming Gaussian distributions around values of [Fe/H] and $V_{\phi}$  and a dispersion according to their estimated errors.
As shown in Figure \ref{figure7}, there exist a clear gradient of
rotational velocity with metallicity [Fe/H] for both the likely
thin-disk and the likely thick-disk sub-sample at different distance
from the Galactic plane. A strong gradient of about $28.1$ to $39.6$
km s$^{-1}$ dex$^{-1}$ for the likely thick-disk sub-sample and a
clear gradient of about $-6.5$ to $-9.5$ km s$^{-1}$ dex$^{-1}$ for
the likely thin-disk sub-sample are obtained in the different
distance slice $|Z|$.  \citet{Lee11} also reported similar gradients
for their likely thin- and thick-disk sub-sample separated base on
[$\alpha$/Fe] ratio versus [Fe/H] derived from spectra of G-type
dwarfs from the SEGUE survey. The gradient for the thick-disk
population agrees with the claim of \citet{Spagna10}, who derived a
similar gradient of $40$ to $50$ km s$^{-1}$ dex$^{-1}$, amongst
their thick disk stars located between 1 and 3 kpc from the Galactic
plane and with metallicity $-1.0 <$ [Fe/H] $< -0.5$.

\begin{table}
\begin{center}
\caption{Rotational Properties of the Selected Sample}
\label{table1}
\begin{tabular}{cccc}
\hline
\hline
[Fe/H]  &  {N$_{\rm Stars}$}  & $\langle V_{\phi} \rangle$ &  $\sigma_{V_{\phi}}$

 \\
          (dex)    &                                &          (km~s$^{-1}$)          &        (km~s$^{-1}$)
 \\
\hline
\multicolumn{4}{c}{0.1$<|Z|$ $<$ 0.5 kpc} \\
\hline
$-1.2$ to $-0.9$ & 142   & 141.4 $\pm$ 5.6    & 66.4 $\pm$ 4.0\\
$-0.9$ to $-0.6$ & 1392   & 172.1 $\pm$ 1.3    & 50.7 $\pm$ 1.0 \\
$-0.6$ to $-0.3$ & 6285 & 198.4 $\pm$ 0.5    & 36.1 $\pm$ 0.3\\
$-0.3$ to $0$   & 7702 & 204.3 $\pm$ 0.3    & 29.1 $\pm$ 0.2\\
$0$ to $0.3$     & 3478 & 200.2 $\pm$ 0.5    & 27.9 $\pm$ 0.3 \\
$0.3$ to $0.6$   & 275   & 198.5 $\pm$ 1.5    & 25.6 $\pm$ 1.1  \\
\hline

\multicolumn{4}{c}{0.5$<$ $|Z|$ $<$ 1.0 kpc} \\
\hline
$-1.2$ to $-0.9$ & 565   & 133.8 $\pm$ 3.2    & 76.2 $\pm$ 2.3\\
$-0.9$ to $-0.6$ & 7055   & 167.5 $\pm$ 0.6    & 50.6 $\pm$ 0.4 \\
$-0.6$ to $-0.3$ & 25526 & 191.6 $\pm$ 0.2    & 39.4 $\pm$ 0.2\\
$-0.3$ to $0$   & 26643 & 198.5 $\pm$ 0.2    & 32.0 $\pm$ 0.2\\
$0$ to $0.3$     & 11566 & 196.5 $\pm$ 0.3    & 30.3 $\pm$ 0.2 \\
$0.3$ to $0.6$   & 911   & 191.0 $\pm$ 1.1    & 31.8 $\pm$ 0.7  \\
\hline

\multicolumn{4}{c}{1.0$<$ $|Z|$ $<$ 3.0 kpc} \\
\hline
$-1.2$ to $-0.9$ & 439   & 133.4 $\pm$ 3.7    & 77.7 $\pm$ 2.6\\
$-0.9$ to $-0.6$ & 6059   & 158.7 $\pm$ 0.7    & 50.1 $\pm$ 0.5 \\
$-0.6$ to $-0.3$ & 16759 & 180.5 $\pm$ 0.4    & 47.7 $\pm$ 0.3\\
$-0.3$ to $0$   & 11023 & 188.8 $\pm$ 0.4    & 40.9 $\pm$ 0.5\\
$0$ to $0.3$     & 3840 & 184.3 $\pm$ 0.7    & 44.3 $\pm$ 0.5 \\
$0.3$ to $0.6$   & 383   & 174.8 $\pm$ 2.6    & 50.5 $\pm$ 1.8  \\

\hline
\end{tabular}
\end{center}
\end{table}

\begin{figure}
\includegraphics[width=1.0\hsize]{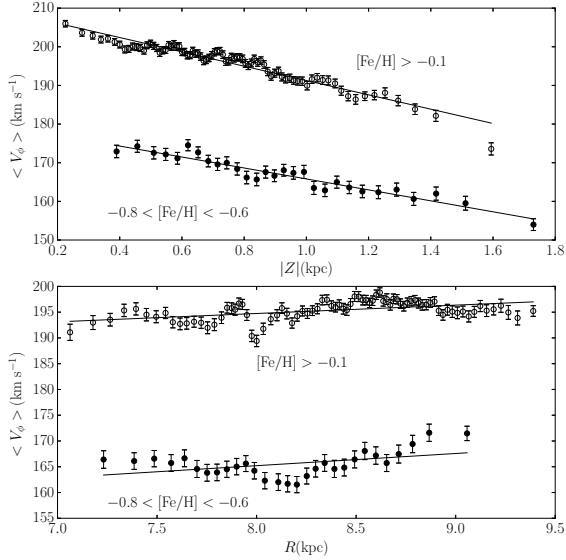}
\caption{The mean rotational velocity with distance from the
Galactic plane $|Z|$ (top panel) and with Galactocentric radius $R$
(bottom panel) for two different metallicity intervals.}
\label{figure8}
\end{figure}

Moreover, Figure \ref{figure6} also indicates the rotational velocity becomes
slow with increased distance from Galactic plane $|Z|$. In order to
explore this behavior more detailed, we select stars in
the high-metallicity range [Fe/H] $> -0.1$ and
intermediate-metallicity range $-0.8 <$ [Fe/H] $< -0.6$, and derive
the mean rotational velocity as a function of $|Z|$ and $R$.  We
derive the mean rotation velocity by passing a box of 1000 stars,
with an overlap of 600 stars per bin through the data.  The results
are given in Figure
\ref{figure8}. We obtain the gradients of mean
rotational velocity $V_\phi$ with distance from the Galactic plane
are $-18.5~\rm km~s^{-1}~kpc^{-1}$ and $-14.2~ \rm km ~
\rm~s^{-1}~kpc^{-1}$ for the high-metallicity and
intermediate-metallicity sub-sample, respectively.  The gradient of
intermediate-metallicity sub-sample is smaller than that obtained by
\citet{Carollo10}, $-36~\rm km~s^{-1}~kpc^{-1}$, base on same
metallicity range $-0.8 <$ [Fe/H] $< -0.6$ at 0 $< |Z|<$ 4 kpc, and
larger than that obtained by \citet{Lee11}, $-9.4~\rm
km~s^{-1}~kpc^{-1}$ for thick-disk stars. The bottom panel of Figure
\ref{figure8} indicates only a negligible rotational velocity
gradient with the Galactocentric radius $R$ for both sub-samples
(only $2.4~\rm km~s^{-1}~kpc^{-1}$ and $1.6~ \rm km ~ \rm~s^{-1} ~
kpc^{-1}$), which is consistent with a flat rotation curve in the
solar neighborhood.

\subsection{Correlation of velocity dispersion with metallicity and $|Z|$}

\begin{figure*}
	\includegraphics[width=1.0\hsize]{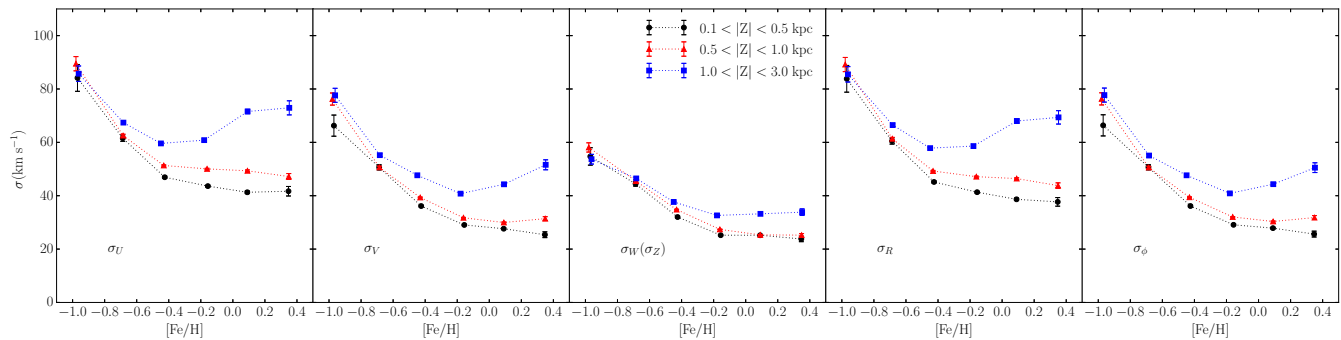}
	\caption{Distribution of the velocity dispersion ($\sigma_{U},
		\sigma_{V}, \sigma_{W}(\sigma_{Z}), \sigma_{R}, \sigma_{\phi}$) vs. [Fe/H] for the star in different $|Z|$
		intervals. The filled circles, triangles and squares
		denote the stars at $0.1<|$Z$|<$ 0.5 kpc, 0.5 $<|$Z$| <$ 1 kpc, and 1 $<
		|$Z$|<$ 3 kpc, respectively. The bins in metallicity are same with Table \ref{table1}.}
	\label{figure9}
\end{figure*}

In this section, we investigate the correlation of velocity dispersion with metallicity and vertical distance from the Galactic plane. Figure \ref{figure9} shows the observed velocity dispersion
($\sigma_{U}, \sigma_{V}, \sigma_{W}(\sigma_{Z}), \sigma_{R}, \sigma_{\phi}$) as a function of metallicity
[Fe/H] at different $|Z|$ intervals.
We first examine the properties of $\sigma_{U}, \sigma_{V}$ and $ \sigma_{W}$.
As shown in Figure \ref{figure9},  the  behavior of the velocity dispersion $\sigma_{U}, \sigma_{V}$ and $ \sigma_{W}$ versus metallicity [Fe/H]
is particularly different at $-1.2 <$ [Fe/H] $< -0.2$ and [Fe/H] $>
-0.2$. In the first metallicity interval, the velocity
dispersions have a relatively larger negative gradient with increased
metallicity [Fe/H] in any distance $|Z|$. Especially for [Fe/H] $<-0.6$, the trends of the velocity dispersion with metallicity for the three velocity
components are consistent with the results from 1203
solar-neighborhood metal-poor stars with
[Fe/H]$<-0.6$ in \citet{Chiba00}. In the second metallicity
range, most stars belong to the thin-disk population, the gradient
is negligible at 0.1 $<$ $|Z|$ $<$ 0.5 kpc and 0.5 $<$ $|Z|$
$<$ 1 kpc, while the velocity dispersions $\sigma_{U}$ and $\sigma_{V}$ begin to increase
with increased metallicity [Fe/H] at 1 $<|$Z$|<$ 3 kpc. The trend for metal-rich star at 1 $<|$Z$|<$ 3 kpc is similar with Figure \ref{figure5},  which reveal those stars possibly belong to thick-disk population because of the larger velocity dispersion and eccentricity. We also notice that the behaviors of velocity dispersions $\sigma_{R},$ and $\sigma_{\phi}$ resemble $\sigma_{U},$ and $\sigma_{\textsc{V}}$ with metallicity [Fe/H] and $|Z|$. This is a natural consequence of our sample far away from Galactic center and having relative small-scale spatial distribution.

\begin{table*}
	\centering
	\caption{
		\label{table2}The Gradients of Velocity Dispersions with $|Z|$}
    \begin{tabular}{cccccc}
		\hline \hline
		[Fe/H]
		& $\partial\sigma_{U}/\partial Z$
		& $\partial\sigma_{V}/\partial Z$
		& $\partial\sigma_{\textsc{W}}/\partial Z(\partial\sigma_{Z}/\partial Z)$
		& $\partial\sigma_{R}/\partial Z$
		& $\partial\sigma_{\phi}/\partial Z$
		\\
		(dex)
		& \multicolumn{5}{c}{------------------------------~~(km s$^{-1}$ kpc$^{-1}$)~~------------------------------}
		\\
		\hline
	    $-1.2$ to $-0.6$ & 5.7 $\pm$ 0.8 & 5.4 $\pm$ 0.7 & 1.6 $\pm$ 0.5  & 5.9 $\pm$ 0.8 & 5.3 $\pm$ 0.7   \\
        $-0.6$ to $-0.1$ & 14.5 $\pm$ 0.3 & 11.9 $\pm$ 0.2 & 7.2 $\pm$ 0.2  & 14.6 $\pm$ 0.3 & 12.0 $\pm$ 0.2  \\
        $-0.1$ to $0.6$	 & 24.1 $\pm$ 0.6 & 13.3 $\pm$ 0.4& 6.5 $\pm$ 0.3  & 23.7 $\pm$ 0.5 & 13.2 $\pm$ 0.4  \\
		\hline
	\end{tabular}
\end{table*}

In addition, Figure \ref{figure9} clearly show an increase of velocity dispersions with increased $|Z|$, especially for the metal-rich stars. We derive the gradient of dispersions with distance $|Z|$ at metallicity ranges $-1.2 <$ [Fe/H] $< -0.6$, $-0.6 <$ [Fe/H] $< -0.1$, and $-0.1 <$ [Fe/H] $< 0.6$.  The values are also listed in Table \ref{table2}. The gradients are obtained by performing least-square fits to the value of dispersions which are derived by passing a box of 800 stars, with an overlap of 300 stars per bin, through the data of each velocity component. As shown in Table \ref{table2}, metal-rich stars have larger gradient of dispersions with $Z$. In metallicity range $-1.2 <$ [Fe/H] $< -0.6$, most stars
belong to the thick-disk population and shows a smaller gradient (5.7, 5.4,
and 1.6 $\rm km~ s^{-1}~kpc^{-1}$) of $\sigma_{U}, \sigma_{V}$, and
$\sigma_{W}$ with distance $|Z|$. In contrast, \citet{MoniBidin10} derived a
vertical gradient of $\sigma_{U}, \sigma_{V}$, and $\sigma_{W}$ of
6.2, 4.5, and 2.8 $\rm km~ s^{-1}~kpc^{-1}$ for thick-disk stars at
2 $<|Z|<$ 4.5 kpc, and \citet{MoniBidin12} also derived a similar gradient of
6.3, 4.1, and 2.4 $\rm km~ s^{-1}~kpc^{-1}$ for stars at 1.5 $<|Z|<$
4.5 kpc.

\section{Discussions}
\subsection{Correlations between Rotational Velocity and Metallicity}
The radial migration models \citep{Sellwood02,Roskar08a,Schonrich09,Sales09,Minchev10}
suggest that the energy and angular momentum changes occur from
interactions with transient spiral arms, which move stars at the
corotation resonance inward or outward in radius while preserving
their nearly circular orbits \citep{Loebman11}.
\citet{Loebman11} employed a $N$-body simulation with radial migration and
found a gradient of about $-24.8$ km s$^{-1}$
dex$^{-1}$ with [Fe/H] for the rotational velocities of their younger stars (identified with thin-disk component), the gradient was more larger when restricting the metallicity with $-0.2<$ [Fe/H] $<0.4$. The result is qualitative agreements with our likely thin-disk component (identified by kinematic criteria).
Loebman et al. (2011) did't find significant gradient (only $1.4$ km s$^{-1}$ dex$^{-1}$) for their older stars ($>$7 Gyr, which generally matched the observed properties of the thick-disk component). This result is disagreement with our observed large positive gradient of rotational velocity with [Fe/H] for likely thick-disk component. However, the pure N-body models of \citet{Curir12} predicted a lager positive slope ($\sim 60$ km s$^{-1}$
dex$^{-1}$) of $V_{\phi}$ with [Fe/H] for their thick-disk population which are at $1.5$ kpc $<|$Z$|<$ $2.0$ kpc and persist up to $6$ Gyr. Their simulation contained radial migration and heating processes of stars from the inner region of the disk and assumed an initial radial chemical gradient as that suggested by \citet{Spitoni11}. As a result, the correlation between $V_{\phi}$ and metallicity [Fe/H] reveal that the radial migration appears to play an important role in the formation of thin-disk population, but it can't reach a firm conclusion about the effect of radial migration in the formation of thick-disk population.

\subsection{Distribution of Orbital Eccentricities of Thick-Disk Stars}

\begin{figure}
\includegraphics[width=1.0\hsize]{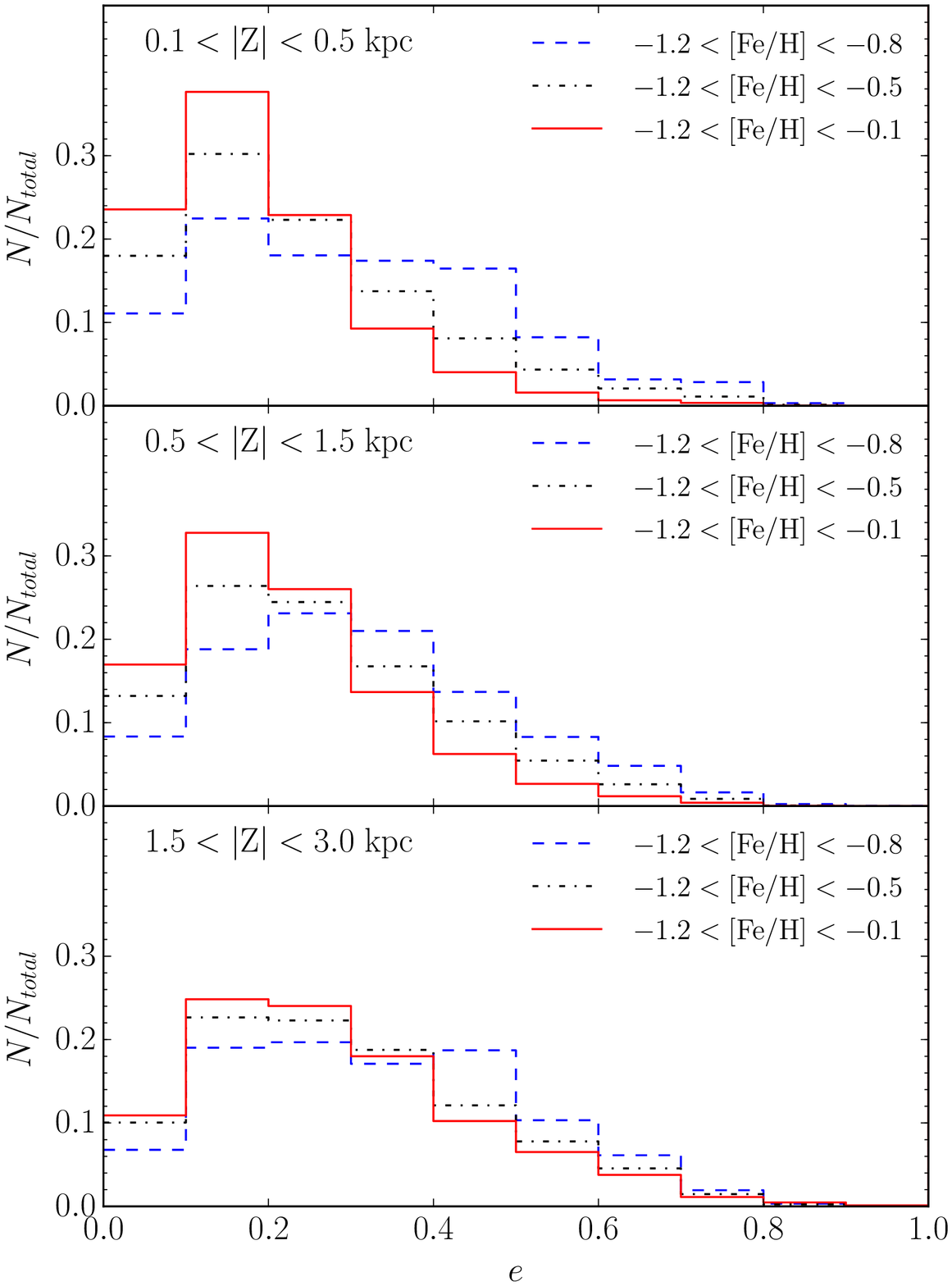}
\caption{Eccentricities distributions for different $|Z|$
and metallicities [Fe/H].}
\label{figure10}
\end{figure}

\citet{Sales09}  demonstrated that the orbital eccentricity distribution of
nearby thick-disk stars may help elucidate the dominant formation mechanism
of the thick-disk and they also show the distribution for different thick-disk
formation models. In this section, we investigate the
eccentricity distribution of observed thick-disk stars and compare the result with
their simulated result.

In order to minimize the contamination from the halo, we only consider stars with
rotational velocity $V_{\phi} >$ 50 km s$^{-1}$, which is consistent with \citet{Sales09}.
Our sample also have eliminated the small fraction of stars with metallicity [Fe/H] $< -1.2$ (\citet{Carollo10}
shows these stars may be chemically attributable to the halo). We then examine the eccentricity
distribution of stars in different range of metallicities and heights from the Galactic plane to estimate
the contamination of the thin-disk, as the thin-disk stars dominate in close distance and metal-rich.
The result is shown in Figure \ref{figure10}. 
It shows that these eccentricities distributions in different $|Z|$
and [Fe/H]  ranges look similar, with a peak at low eccentricity and a
pronounced asymmetric tail toward high eccentricities ($e \sim 0.8$). The thick-disk should dominate at
larger distance ($|$Z$| > 1$kpc) from Galactic plane.
Comparison of the three panels in Figure \ref{figure10} shows that the relative frequency slightly decreases for the high eccentricity and increase for the low eccentricity at smaller $|Z|$ distance where the thin-disk stars should dominate.  Considering more metal-rich stars or more contamination from thin-disk, the same trend of the relative frequency exists.   
We note that these changes are not significant,
and it is robust that eccentricity distribution of thick-disk stars has a peak at $e \sim 0.2$ and exhibits extended tails of higher eccentricities up to $e \sim 0.8$.

\begin{figure}
\includegraphics[width=1.0\hsize]{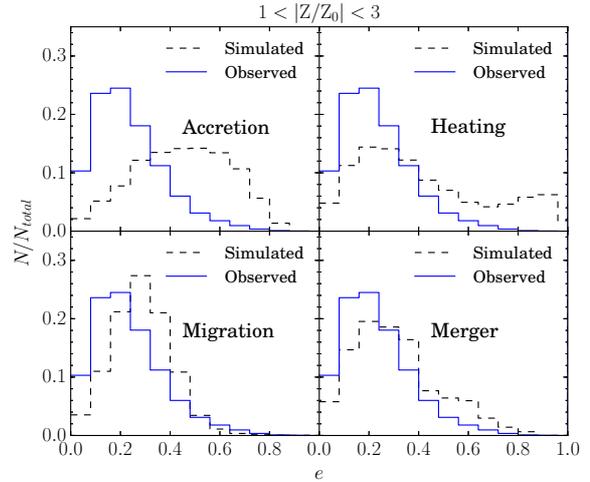}
\caption{Eccentricities distributions for our observed thick-disk stars and model predictions from \citet{Sales09}. The observed data are restricted at $0.8 <|$Z$|< 2.4$ kpc, $-1.2<$ [Fe/H] $<-0.6$ and $V_{\phi} >$ 50 km s$^{-1}$.}
\label{figure11}
\end{figure}

We compare the result with the four published models predictions in \citet{Sales09},
as shown in Figure \ref{figure11}, the four models are accretion model adopted from \citet{Abadi03},
heating model from \citet{Villalobos08} , radial migration model from \citet{Roskar08},  and
 the gas-rich merger model from \citet{Brook04,Brook05}, respectively. The observed thick-disk stars are selected at $0.8 <|$Z$|< 2.4$ kpc
 (taking the scale height Z$_{0}$ of the thick disk as 0.8 kpc) consisting with the range 1 $< |$Z$/$Z$_{0}| <$ 3 in the \citet{Sales09}, $-1.2<$ [Fe/H] $<-0.6$ and also $V_{\phi} >$ 50 km s$^{-1}$. For the accretion model,
 the distribution is symmetric, very broad and having a median eccentricity $e \sim 0.5$, which is not consistent
with our distribution. Although the distributions from heating model has a peak at low eccentricity, it also has secondary
peak at high eccentricity which does not have in our distribution. Figure 3 of \citet{Sales09} shows this secondary peak is mostly occupied by the accreted stars (which retain the initial orbital characteristics of the merging satellite), and \citet{DiMatteo11} found that fewer stars with extreme values and no evidence of their secondary peak around $e \sim 0.8$ in their simulation
with the small satellite mass (1:10 mass ratio). \citet{DiMatteo11} also found that the increase of the stellar orbital
eccentricities in the solar neighborhood with vertical distance,
which can also be found from our Figure \ref{figure10}, can be reproduced if the satellite is accreted onto a direct orbit.
For the migration model, the distribution is more Gaussian like, exhibiting a symmetry about the peak until it gets down to the high-$e$ tail.  But it is not displayed in our distribution, making this particular realization of the migration model somewhat less consistent with our result.
Our result is consistent with the distribution from gas-rich merger model which exhibits an asymmetric peak.

Some studies compared the above models expectations with
observed distributions of orbital eccentricities
for thick-disk stars in the solar neighborhood, e.g.
\citet{Dierickx10}, \citet{Wilson11}, \citet{Lee11}.  Based
on different samples, these studies adopt different models of the Milky Way potential, and all
produce similar eccentricity distributions for the thick-disk stars, namely, a peak at low eccentricity and extended a tail to higher eccentricity. Considering above studies about stellar orbital eccentricities, the favored mechanisms for thick-disk formation are likely to be either (or both) the gas-rich mergers model or the thin-disk heating by minor mergers scenario.

\section{Summary and conclusions }
In this paper, we use
130,043 F/G-type dwarf stars from the LAMOST DR2 data
to investigate kinematics
and metallicity distribution of the Galactic disk.
Our sample comprises stars with  $6.5<R< 9.5$ kpc, $0.1< |Z|< 3$ kpc, log $g
> 3.5$, $-1.2<$[Fe/H]$<0.6$, and $\rm S/N >$ 15.
It shows that our sample stars could be mainly contributed from the
disk system.

In the intermediate-metallicity range $-1.2<$ [Fe/H] $<-0.5$, the orbital eccentricity generally decreases with increased metallicity, and there is a relatively larger gradient. In the metal-rich range $-0.2<$ [Fe/H] $<0.6$, there is little or no correlation between orbital eccentricity and metallicity [Fe/H] for stars at $0.1< |Z|< 0.5$ kpc and $0.5< |Z|< 1.0$ kpc, however, for stars farther from the Galactic plane ($1.0<|Z|<3.0$ kpc), the eccentricity has a little increase with increasing metallicity.  Those trends are also found in the correlation between $\sigma_{U}$ and $\sigma_{V}$ with metallicity.
Moreover, the observed thick-disk stellar orbital eccentricity distribution peaks at low eccentricity ($e \sim 0.2$) and extends to a high eccentricity ($e \sim 0.8$). We then compare this result with four thick-disk formation models, and it appears that the observed distribution is consistent with the gas-rich merger model and against with accretion model.

 We examine the rotational velocity with $|Z|$ and R at the high-metallicity ([Fe/H]$>-0.1$)
and intermediate-metallicity range($-0.8 <$[Fe/H]$<-0.6$). It
show that there exists a clear gradient of $V_{\phi}$ with $|Z|$, and there is only a negligible rotational
velocity gradient with the Galactocentric radius $R$ for both
metallicity range.  In addition, the rotation velocity increases with increased metallicity [Fe/H]
in the range $-1.2<$ Fe/H]$<-0.2$ and has a peak at
[Fe/H]$\sim-0.2$, then it becomes a slightly decrease when
[Fe/H]$>-0.2$. After separating the sample stars into likely thin-disk and thick-disk sub-sample, we find that there
exits a negative gradient of rotation velocity $V_{\phi}$ with metallicity [Fe/H] for the likely thin-disk sub-sample, and the thick-disk sub-sample exhibit a larger positive gradient of rotation velocity with metallicity.
 The gradient for the likely thin-disk sub-sample
qualitatively agrees with the predictions of the radial migration models \citep{Loebman11}.
We consider the radial migration of stars appears to have influenced on the thin-disk formation.
 The gradient for the thick-disk sub-sample is consistent with the result of pure N-body models of \citet{Curir12}.
However, to detailed quantitative comparisons with these
observation results, it need to constructed more physically
realistic models and simulations.

\section*{Acknowledgements}

We especially thank the referee for his/her insightful comments and suggestions
which have improved the paper significantly.
This work was supported by joint fund of Astronomy of
the National Natural Science Foundation of China and the Chinese
Academy of Science, under Grants U1231113.  This work was also by
supported by the Special funds of cooperation between the Institute
and the University of the Chinese Academy of Sciences. In addition,
this work was supported by the National Natural Foundation of China
(NSFC, No.11373033, No.11373035), and by the National Basic Research
Program of China (973 Program) (No. 2014CB845702, No.2014CB845704,
No.2013CB834902). Guoshoujing Telescope (the Large Sky Area Multi-Object Fiber Spectroscopic Telescope LAMOST)
is a National Major Scientific Project built by the Chinese Academy of Sciences.
Funding for the project has been provided by the National Development and Reform Commission.
LAMOST is operated and managed by the National Astronomical Observatories, Chinese Academy of Sciences.


\bibliographystyle{mnras}
\bibliography{bibjyj}

\begin{thebibliography}{}
\makeatletter
\relax
\def\mn@urlcharsother{\let\do\@makeother \do\$\do\&\do\#\do\^\do\_\do\%\do\~}
\def\mn@doi{\begingroup\mn@urlcharsother \@ifnextchar [ {\mn@doi@}
  {\mn@doi@[]}}
\def\mn@doi@[#1]#2{\def\@tempa{#1}\ifx\@tempa\@empty \href
  {http://dx.doi.org/#2} {doi:#2}\else \href {http://dx.doi.org/#2} {#1}\fi
  \endgroup}
\def\mn@eprint#1#2{\mn@eprint@#1:#2::\@nil}
\def\mn@eprint@arXiv#1{\href {http://arxiv.org/abs/#1} {{\tt arXiv:#1}}}
\def\mn@eprint@dblp#1{\href {http://dblp.uni-trier.de/rec/bibtex/#1.xml}
  {dblp:#1}}
\def\mn@eprint@#1:#2:#3:#4\@nil{\def\@tempa {#1}\def\@tempb {#2}\def\@tempc
  {#3}\ifx \@tempc \@empty \let \@tempc \@tempb \let \@tempb \@tempa \fi \ifx
  \@tempb \@empty \def\@tempb {arXiv}\fi \@ifundefined
  {mn@eprint@\@tempb}{\@tempb:\@tempc}{\expandafter \expandafter \csname
  mn@eprint@\@tempb\endcsname \expandafter{\@tempc}}}

\bibitem[\protect\citeauthoryear{{Abadi}, {Navarro}, {Steinmetz}  \&
  {Eke}}{{Abadi} et~al.}{2003}]{Abadi03}
{Abadi} M.~G.,  {Navarro} J.~F.,  {Steinmetz} M.,   {Eke} V.~R.,  2003, \mn@doi
  [\apj] {10.1086/378316}, \href
  {http://adsabs.harvard.edu/abs/2003ApJ...597...21A} {597, 21}

\bibitem[\protect\citeauthoryear{{Aihara} et~al.,}{{Aihara}
  et~al.}{2011}]{Aihara11}
{Aihara} H.,  et~al., 2011, \mn@doi [\apjs] {10.1088/0067-0049/193/2/29}, \href
  {http://adsabs.harvard.edu/abs/2011ApJS..193...29A} {193, 29}

\bibitem[\protect\citeauthoryear{{Beers} et~al.,}{{Beers}
  et~al.}{2006}]{Beers06}
{Beers} T.~C.,  et~al., 2006, \memsai, \href
  {http://adsabs.harvard.edu/abs/2006MmSAI..77.1171B} {77, 1171}

\bibitem[\protect\citeauthoryear{{Bensby}, {Feltzing}  \&
  {Lundstr{\"o}m}}{{Bensby} et~al.}{2003}]{Bensby03}
{Bensby} T.,  {Feltzing} S.,   {Lundstr{\"o}m} I.,  2003, \mn@doi [\aap]
  {10.1051/0004-6361:20031213}, \href
  {http://adsabs.harvard.edu/abs/2003A%26A...410..527B} {410, 527}

\bibitem[\protect\citeauthoryear{{Bensby}, {Zenn}, {Oey}  \&
  {Feltzing}}{{Bensby} et~al.}{2007}]{Bensby07}
{Bensby} T.,  {Zenn} A.~R.,  {Oey} M.~S.,   {Feltzing} S.,  2007, \mn@doi
  [\apjl] {10.1086/519792}, \href
  {http://adsabs.harvard.edu/abs/2007ApJ...663L..13B} {663, L13}

\bibitem[\protect\citeauthoryear{{Bensby}, {Alves-Brito}, {Oey}, {Yong}  \&
  {Mel{\'e}ndez}}{{Bensby} et~al.}{2011}]{Bensby11}
{Bensby} T.,  {Alves-Brito} A.,  {Oey} M.~S.,  {Yong} D.,   {Mel{\'e}ndez} J.,
  2011, \mn@doi [\apjl] {10.1088/2041-8205/735/2/L46}, \href
  {http://adsabs.harvard.edu/abs/2011ApJ...735L..46B} {735, L46}

\bibitem[\protect\citeauthoryear{{Bensby}, {Feltzing}  \& {Oey}}{{Bensby}
  et~al.}{2014}]{Bensby14}
{Bensby} T.,  {Feltzing} S.,   {Oey} M.~S.,  2014, \mn@doi [\aap]
  {10.1051/0004-6361/201322631}, \href
  {http://adsabs.harvard.edu/abs/2014A%26A...562A..71B} {562, A71}

\bibitem[\protect\citeauthoryear{{Binney} et~al.,}{{Binney}
  et~al.}{2014}]{Binney14}
{Binney} J.,  et~al., 2014, \mn@doi [\mnras] {10.1093/mnras/stt2367}, \href
  {http://adsabs.harvard.edu/abs/2014MNRAS.439.1231B} {439, 1231}

\bibitem[\protect\citeauthoryear{{Bond} et~al.,}{{Bond} et~al.}{2010}]{Bond10}
{Bond} N.~A.,  et~al., 2010, \mn@doi [\apj] {10.1088/0004-637X/716/1/1}, \href
  {http://adsabs.harvard.edu/abs/2010ApJ...716....1B} {716, 1}

\bibitem[\protect\citeauthoryear{{Bovy} et~al.,}{{Bovy} et~al.}{2012}]{Bovy12}
{Bovy} J.,  et~al., 2012, \mn@doi [\apj] {10.1088/0004-637X/759/2/131}, \href
  {http://adsabs.harvard.edu/abs/2012ApJ...759..131B} {759, 131}

\bibitem[\protect\citeauthoryear{{Bovy}, {Bird}, {Garc{\'{\i}}a P{\'e}rez},
  {Majewski}, {Nidever}  \& {Zasowski}}{{Bovy} et~al.}{2015}]{Bovy15}
{Bovy} J.,  {Bird} J.~C.,  {Garc{\'{\i}}a P{\'e}rez} A.~E.,  {Majewski} S.~R.,
  {Nidever} D.~L.,   {Zasowski} G.,  2015, \mn@doi [\apj]
  {10.1088/0004-637X/800/2/83}, \href
  {http://adsabs.harvard.edu/abs/2015ApJ...800...83B} {800, 83}

\bibitem[\protect\citeauthoryear{{Brook}, {Kawata}, {Gibson}  \&
  {Freeman}}{{Brook} et~al.}{2004}]{Brook04}
{Brook} C.~B.,  {Kawata} D.,  {Gibson} B.~K.,   {Freeman} K.~C.,  2004, \mn@doi
  [\apj] {10.1086/422709}, \href
  {http://adsabs.harvard.edu/abs/2004ApJ...612..894B} {612, 894}

\bibitem[\protect\citeauthoryear{{Brook}, {Gibson}, {Martel}  \&
  {Kawata}}{{Brook} et~al.}{2005}]{Brook05}
{Brook} C.~B.,  {Gibson} B.~K.,  {Martel} H.,   {Kawata} D.,  2005, \mn@doi
  [\apj] {10.1086/431924}, \href
  {http://adsabs.harvard.edu/abs/2005ApJ...630..298B} {630, 298}

\bibitem[\protect\citeauthoryear{{Brook}, {Richard}, {Kawata}, {Martel}  \&
  {Gibson}}{{Brook} et~al.}{2007}]{Brook07}
{Brook} C.,  {Richard} S.,  {Kawata} D.,  {Martel} H.,   {Gibson} B.~K.,  2007,
  \mn@doi [\apj] {10.1086/511056}, \href
  {http://adsabs.harvard.edu/abs/2007ApJ...658...60B} {658, 60}

\bibitem[\protect\citeauthoryear{{Carollo} et~al.,}{{Carollo}
  et~al.}{2010}]{Carollo10}
{Carollo} D.,  et~al., 2010, \mn@doi [\apj] {10.1088/0004-637X/712/1/692},
  \href {http://adsabs.harvard.edu/abs/2010ApJ...712..692C} {712, 692}

\bibitem[\protect\citeauthoryear{{Casetti-Dinescu}, {Girard}, {Korchagin}  \&
  {van Altena}}{{Casetti-Dinescu} et~al.}{2011}]{CasettiDinescu11}
{Casetti-Dinescu} D.~I.,  {Girard} T.~M.,  {Korchagin} V.~I.,   {van Altena}
  W.~F.,  2011, \mn@doi [\apj] {10.1088/0004-637X/728/1/7}, \href
  {http://adsabs.harvard.edu/abs/2011ApJ...728....7C} {728, 7}

\bibitem[\protect\citeauthoryear{{Cheng} et~al.,}{{Cheng}
  et~al.}{2012}]{Cheng12}
{Cheng} J.~Y.,  et~al., 2012, \mn@doi [\apj] {10.1088/0004-637X/752/1/51},
  \href {http://adsabs.harvard.edu/abs/2012ApJ...752...51C} {752, 51}

\bibitem[\protect\citeauthoryear{{Chiba} \& {Beers}}{{Chiba} \&
  {Beers}}{2000}]{Chiba00}
{Chiba} M.,  {Beers} T.~C.,  2000, \mn@doi [\aj] {10.1086/301409}, \href
  {http://adsabs.harvard.edu/abs/2000AJ....119.2843C} {119, 2843}

\bibitem[\protect\citeauthoryear{{Cui} et~al.,}{{Cui} et~al.}{2012}]{Cui12}
{Cui} X.-Q.,  et~al., 2012, \mn@doi [RAA] {10.1088/1674-4527/12/9/003}, \href
  {http://adsabs.harvard.edu/abs/2012RAA....12.1197C} {12, 1197}

\bibitem[\protect\citeauthoryear{{Curir}, {Lattanzi}, {Spagna}, {Matteucci},
  {Murante}, {Re Fiorentin}  \& {Spitoni}}{{Curir} et~al.}{2012}]{Curir12}
{Curir} A.,  {Lattanzi} M.~G.,  {Spagna} A.,  {Matteucci} F.,  {Murante} G.,
  {Re Fiorentin} P.,   {Spitoni} E.,  2012, \mn@doi [\aap]
  {10.1051/0004-6361/201118558}, \href
  {http://adsabs.harvard.edu/abs/2012A%26A...545A.133C} {545, A133}

\bibitem[\protect\citeauthoryear{{Deng} et~al.,}{{Deng} et~al.}{2012}]{Deng12}
{Deng} L.-C.,  et~al., 2012, \mn@doi [RAA] {10.1088/1674-4527/12/7/003}, \href
  {http://adsabs.harvard.edu/abs/2012RAA....12..735D} {12, 735}

\bibitem[\protect\citeauthoryear{{Di Matteo}, {Lehnert}, {Qu}  \& {van
  Driel}}{{Di Matteo} et~al.}{2011}]{DiMatteo11}
{Di Matteo} P.,  {Lehnert} M.~D.,  {Qu} Y.,   {van Driel} W.,  2011, \mn@doi
  [\aap] {10.1051/0004-6361/201015822}, \href
  {http://adsabs.harvard.edu/abs/2011A%26A...525L...3D} {525, L3}

\bibitem[\protect\citeauthoryear{{Dierickx}, {Klement}, {Rix}  \&
  {Liu}}{{Dierickx} et~al.}{2010}]{Dierickx10}
{Dierickx} M.,  {Klement} R.,  {Rix} H.-W.,   {Liu} C.,  2010, \mn@doi [\apjl]
  {10.1088/2041-8205/725/2/L186}, \href
  {http://adsabs.harvard.edu/abs/2010ApJ...725L.186D} {725, L186}

\bibitem[\protect\citeauthoryear{{Du} et~al.,}{{Du} et~al.}{2003}]{Du03}
{Du} C.,  et~al., 2003, \mn@doi [\aap] {10.1051/0004-6361:20030532}, \href
  {http://adsabs.harvard.edu/abs/2003A%26A...407..541D} {407, 541}

\bibitem[\protect\citeauthoryear{{Du}, {Ma}, {Wu}  \& {Zhou}}{{Du}
  et~al.}{2006}]{Du06}
{Du} C.,  {Ma} J.,  {Wu} Z.,   {Zhou} X.,  2006, \mn@doi [\mnras]
  {10.1111/j.1365-2966.2006.10940.x}, \href
  {http://adsabs.harvard.edu/abs/2006MNRAS.372.1304D} {372, 1304}

\bibitem[\protect\citeauthoryear{{Eisenstein} et~al.,}{{Eisenstein}
  et~al.}{2011}]{Eisenstein11}
{Eisenstein} D.~J.,  et~al., 2011, \mn@doi [\aj] {10.1088/0004-6256/142/3/72},
  \href {http://adsabs.harvard.edu/abs/2011AJ....142...72E} {142, 72}

\bibitem[\protect\citeauthoryear{{Feast} \& {Whitelock}}{{Feast} \&
  {Whitelock}}{1997}]{Feast97}
{Feast} M.~W.,  {Whitelock} P.~A.,  1997, in Hipparcos - Venice '97. pp
  625--628

\bibitem[\protect\citeauthoryear{{Freeman}}{{Freeman}}{1987}]{Freeman87}
{Freeman} K.~C.,  1987, \mn@doi [\araa] {10.1146/annurev.aa.25.090187.003131},
  \href {http://adsabs.harvard.edu/abs/1987ARA%26A..25..603F} {25, 603}

\bibitem[\protect\citeauthoryear{{Freeman} \& {Bland-Hawthorn}}{{Freeman} \&
  {Bland-Hawthorn}}{2002}]{Freeman02}
{Freeman} K.,  {Bland-Hawthorn} J.,  2002, \mn@doi [\araa]
  {10.1146/annurev.astro.40.060401.093840}, \href
  {http://adsabs.harvard.edu/abs/2002ARA%26A..40..487F} {40, 487}

\bibitem[\protect\citeauthoryear{{Fuhrmann}}{{Fuhrmann}}{1998}]{Fuhrmann98}
{Fuhrmann} K.,  1998, \aap, \href
  {http://adsabs.harvard.edu/abs/1998A%26A...338..161F} {338, 161}

\bibitem[\protect\citeauthoryear{{Fuhrmann}}{{Fuhrmann}}{2008}]{Fuhrmann08}
{Fuhrmann} K.,  2008, \mn@doi [\mnras] {10.1111/j.1365-2966.2007.12671.x},
  \href {http://adsabs.harvard.edu/abs/2008MNRAS.384..173F} {384, 173}

\bibitem[\protect\citeauthoryear{{Gilmore} \& {Reid}}{{Gilmore} \&
  {Reid}}{1983}]{Gilmore83}
{Gilmore} G.,  {Reid} N.,  1983, \mn@doi [\mnras] {10.1093/mnras/202.4.1025},
  \href {http://adsabs.harvard.edu/abs/1983MNRAS.202.1025G} {202, 1025}

\bibitem[\protect\citeauthoryear{{Gilmore} \& {Wyse}}{{Gilmore} \&
  {Wyse}}{1985}]{Gilmore85}
{Gilmore} G.,  {Wyse} R.~F.~G.,  1985, \mn@doi [\aj] {10.1086/113907}, \href
  {http://adsabs.harvard.edu/abs/1985AJ.....90.2015G} {90, 2015}

\bibitem[\protect\citeauthoryear{{Gratton}, {Carretta}, {Matteucci}  \&
  {Sneden}}{{Gratton} et~al.}{1996}]{Gratton96}
{Gratton} R.,  {Carretta} E.,  {Matteucci} F.,   {Sneden} C.,  1996, in
  Formation of the Galactic Halo...Inside and Out. p.~307

\bibitem[\protect\citeauthoryear{{Gunn}, {Knapp}  \& {Tremaine}}{{Gunn}
  et~al.}{1979}]{Gunn79}
{Gunn} J.~E.,  {Knapp} G.~R.,   {Tremaine} S.~D.,  1979, \mn@doi [\aj]
  {10.1086/112525}, \href {http://adsabs.harvard.edu/abs/1979AJ.....84.1181G}
  {84, 1181}

\bibitem[\protect\citeauthoryear{{Hayden} et~al.,}{{Hayden}
  et~al.}{2015}]{Hayden15}
{Hayden} M.~R.,  et~al., 2015, \mn@doi [\apj] {10.1088/0004-637X/808/2/132},
  \href {http://adsabs.harvard.edu/abs/2015ApJ...808..132H} {808, 132}

\bibitem[\protect\citeauthoryear{{Ivezi{\'c}} et~al.,}{{Ivezi{\'c}}
  et~al.}{2008}]{Ivezic08}
{Ivezi{\'c}} {\v Z}.,  et~al., 2008, \mn@doi [\apj] {10.1086/589678}, \href
  {http://adsabs.harvard.edu/abs/2008ApJ...684..287I} {684, 287}

\bibitem[\protect\citeauthoryear{{Jia} et~al.,}{{Jia} et~al.}{2014}]{Jia14}
{Jia} Y.,  et~al., 2014, \mn@doi [\mnras] {10.1093/mnras/stu469}, \href
  {\left(} {441, 503}

\bibitem[\protect\citeauthoryear{{Juri{\'c}} et~al.,}{{Juri{\'c}}
  et~al.}{2008}]{Juric08}
{Juri{\'c}} M.,  et~al., 2008, \mn@doi [\apj] {10.1086/523619}, \href
  {http://adsabs.harvard.edu/abs/2008ApJ...673..864J} {673, 864}

\bibitem[\protect\citeauthoryear{{Kordopatis} et~al.,}{{Kordopatis}
  et~al.}{2015}]{Kordopatis15}
{Kordopatis} G.,  et~al., 2015, \mn@doi [\aap] {10.1051/0004-6361/201526258},
  \href {http://adsabs.harvard.edu/abs/2015A%26A...582A.122K} {582, A122}

\bibitem[\protect\citeauthoryear{{Lee} et~al.,}{{Lee} et~al.}{2011}]{Lee11}
{Lee} Y.~S.,  et~al., 2011, \mn@doi [\apj] {10.1088/0004-637X/738/2/187}, \href
  {http://adsabs.harvard.edu/abs/2011ApJ...738..187L} {738, 187}

\bibitem[\protect\citeauthoryear{{Loebman}, {Ro{\v s}kar}, {Debattista},
  {Ivezi{\'c}}, {Quinn}  \& {Wadsley}}{{Loebman} et~al.}{2011}]{Loebman11}
{Loebman} S.~R.,  {Ro{\v s}kar} R.,  {Debattista} V.~P.,  {Ivezi{\'c}} {\v Z}.,
   {Quinn} T.~R.,   {Wadsley} J.,  2011, \mn@doi [\apj]
  {10.1088/0004-637X/737/1/8}, \href
  {http://adsabs.harvard.edu/abs/2011ApJ...737....8L} {737, 8}

\bibitem[\protect\citeauthoryear{{Luo} et~al.,}{{Luo} et~al.}{2012}]{Luo12}
{Luo} A.-L.,  et~al., 2012, \mn@doi [RAA] {10.1088/1674-4527/12/9/004}, \href
  {http://adsabs.harvard.edu/abs/2012RAA....12.1243L} {12, 1243}

\bibitem[\protect\citeauthoryear{{Majewski}}{{Majewski}}{1993}]{majewski93}
{Majewski} S.~R.,  1993, \mn@doi [\araa] {10.1146/annurev.aa.31.090193.003043},
  \href {http://adsabs.harvard.edu/abs/1993ARA%26A..31..575M} {31, 575}

\bibitem[\protect\citeauthoryear{{Minchev} \& {Famaey}}{{Minchev} \&
  {Famaey}}{2010}]{Minchev10}
{Minchev} I.,  {Famaey} B.,  2010, \mn@doi [\apj]
  {10.1088/0004-637X/722/1/112}, \href
  {http://adsabs.harvard.edu/abs/2010ApJ...722..112M} {722, 112}

\bibitem[\protect\citeauthoryear{{Miyamoto} \& {Nagai}}{{Miyamoto} \&
  {Nagai}}{1975}]{Miyamoto75}
{Miyamoto} M.,  {Nagai} R.,  1975, \pasj, \href
  {http://adsabs.harvard.edu/abs/1975PASJ...27..533M} {27, 533}

\bibitem[\protect\citeauthoryear{{Moni Bidin}, {Carraro}, {M{\'e}ndez}  \& {van
  Altena}}{{Moni Bidin} et~al.}{2010}]{MoniBidin10}
{Moni Bidin} C.,  {Carraro} G.,  {M{\'e}ndez} R.~A.,   {van Altena} W.~F.,
  2010, \mn@doi [\apjl] {10.1088/2041-8205/724/1/L122}, \href
  {http://adsabs.harvard.edu/abs/2010ApJ...724L.122M} {724, L122}

\bibitem[\protect\citeauthoryear{{Moni Bidin}, {Carraro}  \&
  {M{\'e}ndez}}{{Moni Bidin} et~al.}{2012}]{MoniBidin12}
{Moni Bidin} C.,  {Carraro} G.,   {M{\'e}ndez} R.~A.,  2012, \mn@doi [\apj]
  {10.1088/0004-637X/747/2/101}, \href
  {http://adsabs.harvard.edu/abs/2012ApJ...747..101M} {747, 101}

\bibitem[\protect\citeauthoryear{{Moultaka}, {Ilovaisky}, {Prugniel}  \&
  {Soubiran}}{{Moultaka} et~al.}{2004}]{Moultaka04}
{Moultaka} J.,  {Ilovaisky} S.~A.,  {Prugniel} P.,   {Soubiran} C.,  2004,
  \mn@doi [\pasp] {10.1086/422177}, \href
  {http://adsabs.harvard.edu/abs/2004PASP..116..693M} {116, 693}

\bibitem[\protect\citeauthoryear{{Munn} et~al.,}{{Munn} et~al.}{2004}]{Munn04}
{Munn} J.~A.,  et~al., 2004, \mn@doi [\aj] {10.1086/383292}, \href
  {http://adsabs.harvard.edu/abs/2004AJ....127.3034M} {127, 3034}

\bibitem[\protect\citeauthoryear{{Munn} et~al.,}{{Munn} et~al.}{2008}]{Munn08}
{Munn} J.~A.,  et~al., 2008, \mn@doi [\aj] {10.1088/0004-6256/136/2/895}, \href
  {http://adsabs.harvard.edu/abs/2008AJ....136..895M} {136, 895}

\bibitem[\protect\citeauthoryear{{Ness} et~al.,}{{Ness} et~al.}{2016}]{Ness16}
{Ness} M.,  et~al., 2016, \mn@doi [\apj] {10.3847/0004-637X/819/1/2}, \href
  {http://adsabs.harvard.edu/abs/2016ApJ...819....2N} {819, 2}

\bibitem[\protect\citeauthoryear{{Ojha}, {Bienayme}, {Robin}, {Creze}  \&
  {Mohan}}{{Ojha} et~al.}{1996}]{Ojha96}
{Ojha} D.~K.,  {Bienayme} O.,  {Robin} A.~C.,  {Creze} M.,   {Mohan} V.,  1996,
  \aap, \href {http://adsabs.harvard.edu/abs/1996A%26A...311..456O} {311, 456}

\bibitem[\protect\citeauthoryear{{Paczynski}}{{Paczynski}}{1990}]{Paczynski90}
{Paczynski} B.,  1990, \mn@doi [\apj] {10.1086/168257}, \href
  {http://adsabs.harvard.edu/abs/1990ApJ...348..485P} {348, 485}

\bibitem[\protect\citeauthoryear{{Pomp{\'e}ia}, {Barbuy}  \&
  {Grenon}}{{Pomp{\'e}ia} et~al.}{2002}]{Pompeia02}
{Pomp{\'e}ia} L.,  {Barbuy} B.,   {Grenon} M.,  2002, \mn@doi [\apj]
  {10.1086/338111}, \href {http://adsabs.harvard.edu/abs/2002ApJ...566..845P}
  {566, 845}

\bibitem[\protect\citeauthoryear{{Prochaska}, {Naumov}, {Carney}, {McWilliam}
  \& {Wolfe}}{{Prochaska} et~al.}{2000}]{Prochaska00}
{Prochaska} J.~X.,  {Naumov} S.~O.,  {Carney} B.~W.,  {McWilliam} A.,   {Wolfe}
  A.~M.,  2000, \mn@doi [\aj] {10.1086/316818}, \href
  {http://adsabs.harvard.edu/abs/2000AJ....120.2513P} {120, 2513}

\bibitem[\protect\citeauthoryear{{Quinn} \& {Goodman}}{{Quinn} \&
  {Goodman}}{1986}]{Quinn86}
{Quinn} P.~J.,  {Goodman} J.,  1986, \mn@doi [\apj] {10.1086/164619}, \href
  {http://adsabs.harvard.edu/abs/1986ApJ...309..472Q} {309, 472}

\bibitem[\protect\citeauthoryear{{Recio-Blanco} et~al.,}{{Recio-Blanco}
  et~al.}{2014}]{RecioBlanco14}
{Recio-Blanco} A.,  et~al., 2014, \mn@doi [\aap] {10.1051/0004-6361/201322944},
  \href {http://adsabs.harvard.edu/abs/2014A%26A...567A...5R} {567, A5}

\bibitem[\protect\citeauthoryear{{Reddy}, {Lambert}  \& {Allende
  Prieto}}{{Reddy} et~al.}{2006}]{Reddy06}
{Reddy} B.~E.,  {Lambert} D.~L.,   {Allende Prieto} C.,  2006, \mn@doi [\mnras]
  {10.1111/j.1365-2966.2006.10148.x}, \href
  {http://adsabs.harvard.edu/abs/2006MNRAS.367.1329R} {367, 1329}

\bibitem[\protect\citeauthoryear{{Ro{\v s}kar}, {Debattista}, {Stinson},
  {Quinn}, {Kaufmann}  \& {Wadsley}}{{Ro{\v s}kar} et~al.}{2008a}]{Roskar08}
{Ro{\v s}kar} R.,  {Debattista} V.~P.,  {Stinson} G.~S.,  {Quinn} T.~R.,
  {Kaufmann} T.,   {Wadsley} J.,  2008a, \mn@doi [\apjl] {10.1086/586734},
  \href {http://adsabs.harvard.edu/abs/2008ApJ...675L..65R} {675, L65}

\bibitem[\protect\citeauthoryear{{Ro{\v s}kar}, {Debattista}, {Quinn},
  {Stinson}  \& {Wadsley}}{{Ro{\v s}kar} et~al.}{2008b}]{Roskar08a}
{Ro{\v s}kar} R.,  {Debattista} V.~P.,  {Quinn} T.~R.,  {Stinson} G.~S.,
  {Wadsley} J.,  2008b, \mn@doi [\apjl] {10.1086/592231}, \href
  {http://adsabs.harvard.edu/abs/2008ApJ...684L..79R} {684, L79}

\bibitem[\protect\citeauthoryear{{Sales} et~al.,}{{Sales}
  et~al.}{2009}]{Sales09}
{Sales} L.~V.,  et~al., 2009, \mn@doi [\mnras]
  {10.1111/j.1745-3933.2009.00763.x}, \href
  {http://adsabs.harvard.edu/abs/2009MNRAS.400L..61S} {400, L61}

\bibitem[\protect\citeauthoryear{{Sch{\"o}nrich} \& {Binney}}{{Sch{\"o}nrich}
  \& {Binney}}{2009}]{Schonrich09}
{Sch{\"o}nrich} R.,  {Binney} J.,  2009, \mn@doi [\mnras]
  {10.1111/j.1365-2966.2009.14750.x}, \href
  {http://adsabs.harvard.edu/abs/2009MNRAS.396..203S} {396, 203}

\bibitem[\protect\citeauthoryear{{Sch{\"o}nrich}, {Binney}  \&
  {Dehnen}}{{Sch{\"o}nrich} et~al.}{2010}]{Schonrich10}
{Sch{\"o}nrich} R.,  {Binney} J.,   {Dehnen} W.,  2010, \mn@doi [\mnras]
  {10.1111/j.1365-2966.2010.16253.x}, \href
  {http://adsabs.harvard.edu/abs/2010MNRAS.403.1829S} {403, 1829}

\bibitem[\protect\citeauthoryear{{Sellwood} \& {Binney}}{{Sellwood} \&
  {Binney}}{2002}]{Sellwood02}
{Sellwood} J.~A.,  {Binney} J.~J.,  2002, \mn@doi [\mnras]
  {10.1046/j.1365-8711.2002.05806.x}, \href
  {http://adsabs.harvard.edu/abs/2002MNRAS.336..785S} {336, 785}

\bibitem[\protect\citeauthoryear{{Siebert} et~al.,}{{Siebert}
  et~al.}{2008}]{Siebert08}
{Siebert} A.,  et~al., 2008, \mn@doi [\mnras]
  {10.1111/j.1365-2966.2008.13912.x}, \href
  {http://adsabs.harvard.edu/abs/2008MNRAS.391..793S} {391, 793}

\bibitem[\protect\citeauthoryear{{Siebert} et~al.,}{{Siebert}
  et~al.}{2011}]{Siebert11}
{Siebert} A.,  et~al., 2011, \mn@doi [\aj] {10.1088/0004-6256/141/6/187}, \href
  {http://adsabs.harvard.edu/abs/2011AJ....141..187S} {141, 187}

\bibitem[\protect\citeauthoryear{{Smith} et~al.,}{{Smith}
  et~al.}{2009}]{Smith09}
{Smith} M.~C.,  et~al., 2009, \mn@doi [\mnras]
  {10.1111/j.1365-2966.2009.15391.x}, \href
  {http://adsabs.harvard.edu/abs/2009MNRAS.399.1223S} {399, 1223}

\bibitem[\protect\citeauthoryear{{Spagna}, {Lattanzi}, {Re Fiorentin}  \&
  {Smart}}{{Spagna} et~al.}{2010}]{Spagna10}
{Spagna} A.,  {Lattanzi} M.~G.,  {Re Fiorentin} P.,   {Smart} R.~L.,  2010,
  \mn@doi [\aap] {10.1051/0004-6361/200913538}, \href
  {http://adsabs.harvard.edu/abs/2010A%26A...510L...4S} {510, L4}

\bibitem[\protect\citeauthoryear{{Spitoni} \& {Matteucci}}{{Spitoni} \&
  {Matteucci}}{2011}]{Spitoni11}
{Spitoni} E.,  {Matteucci} F.,  2011, \mn@doi [\aap]
  {10.1051/0004-6361/201015749}, \href
  {http://adsabs.harvard.edu/abs/2011A%26A...531A..72S} {531, A72}

\bibitem[\protect\citeauthoryear{{Su} \& {Cui}}{{Su} \& {Cui}}{2004}]{Su04}
{Su} D.-Q.,  {Cui} X.-Q.,  2004, \cjaa, \href
  {http://adsabs.harvard.edu/abs/2004ChJAA...4....1S} {4, 1}

\bibitem[\protect\citeauthoryear{{Tian} et~al.,}{{Tian} et~al.}{2015}]{Tian15}
{Tian} H.-J.,  et~al., 2015, \mn@doi [\apj] {10.1088/0004-637X/809/2/145},
  \href {http://adsabs.harvard.edu/abs/2015ApJ...809..145T} {809, 145}

\bibitem[\protect\citeauthoryear{{Villalobos} \& {Helmi}}{{Villalobos} \&
  {Helmi}}{2008}]{Villalobos08}
{Villalobos} {\'A}.,  {Helmi} A.,  2008, \mn@doi [\mnras]
  {10.1111/j.1365-2966.2008.13979.x}, \href
  {http://adsabs.harvard.edu/abs/2008MNRAS.391.1806V} {391, 1806}

\bibitem[\protect\citeauthoryear{{Villalobos}, {Kazantzidis}  \&
  {Helmi}}{{Villalobos} et~al.}{2010}]{Villalobos10}
{Villalobos} {\'A}.,  {Kazantzidis} S.,   {Helmi} A.,  2010, \mn@doi [\apj]
  {10.1088/0004-637X/718/1/314}, \href
  {http://adsabs.harvard.edu/abs/2010ApJ...718..314V} {718, 314}

\bibitem[\protect\citeauthoryear{{Wang}, {Su}, {Chu}, {Cui}  \& {Wang}}{{Wang}
  et~al.}{1996}]{Wang96}
{Wang} S.-G.,  {Su} D.-Q.,  {Chu} Y.-Q.,  {Cui} X.,   {Wang} Y.-N.,  1996,
  \mn@doi [\ao] {10.1364/AO.35.005155}, \href
  {http://adsabs.harvard.edu/abs/1996ApOpt..35.5155W} {35, 5155}

\bibitem[\protect\citeauthoryear{{Wilson} et~al.,}{{Wilson}
  et~al.}{2011}]{Wilson11}
{Wilson} M.~L.,  et~al., 2011, \mn@doi [\mnras]
  {10.1111/j.1365-2966.2011.18298.x}, \href
  {http://adsabs.harvard.edu/abs/2011MNRAS.413.2235W} {413, 2235}

\bibitem[\protect\citeauthoryear{{Wu} et~al.,}{{Wu} et~al.}{2011}]{Wu11}
{Wu} Y.,  et~al., 2011, \mn@doi [RAA] {10.1088/1674-4527/11/8/006}, \href
  {http://adsabs.harvard.edu/abs/2011RAA....11..924W} {11, 924}

\bibitem[\protect\citeauthoryear{{Yanny} et~al.,}{{Yanny}
  et~al.}{2009}]{Yanny09}
{Yanny} B.,  et~al., 2009, \mn@doi [\aj] {10.1088/0004-6256/137/5/4377}, \href
  {http://adsabs.harvard.edu/abs/2009AJ....137.4377Y} {137, 4377}

\bibitem[\protect\citeauthoryear{{York} et~al.,}{{York} et~al.}{2000}]{York00}
{York} D.~G.,  et~al., 2000, \mn@doi [\aj] {10.1086/301513}, \href
  {http://adsabs.harvard.edu/abs/2000AJ....120.1579Y} {120, 1579}

\bibitem[\protect\citeauthoryear{{Zhao}, {Zhao}, {Chu}, {Jing}  \&
  {Deng}}{{Zhao} et~al.}{2012}]{Zhao12}
{Zhao} G.,  {Zhao} Y.-H.,  {Chu} Y.-Q.,  {Jing} Y.-P.,   {Deng} L.-C.,  2012,
  \mn@doi [RAA] {10.1088/1674-4527/12/7/002}, \href
  {http://adsabs.harvard.edu/abs/2012RAA....12..723Z} {12, 723}

\makeatother
\end{thebibliography}

\bsp    
\label{lastpage}
\end{document}